\documentclass{PoS}

\usepackage{amsfonts}
\usepackage{amsmath}
\usepackage{graphics}
\usepackage{epsfig}
\usepackage{multirow}

\newcommand {\be}{\begin{equation}}
\newcommand {\ee}{\end{equation}}
\newcommand {\ba}{\begin{eqnarray}}
\newcommand {\ea}{\end{eqnarray}}

\newcommand {\tanb}{$\tan\beta~$}
\newcommand {\ra}{\rightarrow}

\newcommand {\sbma}{s_{\beta-\alpha}}
\newcommand {\cbma}{c_{\beta-\alpha}}
\newcommand {\sw}{s_{W}}

\title{Charged Higgs Study through Triple or Double Higgs Production in a General Two Higgs Doublet Model at Future Lepton Collider}

\ShortTitle{2HDM Type II $\&$ Triple Higgs Production at Linear Collider}

\author{\speaker{Ijaz Ahmed}\thanks{corresponding author}\\
        COMSATS Institute of Information Technology (CIIT), Islamabad 44000, Pakistan\\
        National Center for Particle Physics, Institute of Research Management \& Monitoring, University of Malaya, 50603 Kuala Lumpur, Malaysia\\
        E-mail: \email{ijaz.ahmed@cern.ch}}

\author{M. Hashemi\\
        Physics Department and Biruni Observatory, College of Sciences, Shiraz University, Shiraz 71454, Iran\\
        E-mail: \email{hashemi$\_$mj@shirazu.ac.ir}}

\abstract{In this study the charged Higgs signal through triple or double Higgs production in a general two Higgs doublet model (2HDM) is investigated. The main production process is $e^+e^- \rightarrow H^+H^-H^0$ followed by the charged Higgs decay to a pair of $\tau \nu$ and the neutral Higgs decay to $b\bar{b}$. The alternative process $H^+W^-H^0$ is also included as a source of charged Higgs signal in the analysis having comparable cross-section. The focus is on a future $e^+e^-$ linear collider operating at $\sqrt{s}=1.5$ TeV. The final state under consideration ($\tau^+ \tau^- b \bar{b} E^{\textnormal{miss}}_{T}$) is suitable for electroweak background rejection using the $b-$tagging tools. It is shown that although the signal cross section is small, with a reasonable background suppression, high signal significance values are achievable at an integrated luminosity $500 fb^{-1}$ depending on the charged Higgs mass, $\tan\beta~$ and the CP-odd neutral Higgs mass. Finally results are quoted in terms of the signal significance for charged Higgs in the mass range $170<m_{H^{\pm}}<400$ GeV.}

\FullConference{Prospects for Charged Higgs Discovery at Colliders - CHARGED 2014,\\
		16-18 September 2014\\
		Uppsala University, Sweden}

\begin{document}

\section{Introduction}
In a general 2HDM, the Higgs sector consists of two charged Higgs bosons, $H^{\pm}$, two CP-even neutral Higgs bosons, $h^0,~H^0$, and a CP-odd neutral Higgs, $A^0$. The lightest neutral Higgs boson $h^0$, is taken to be SM-like and is the candidate for the signal observed at LHC. 
A charged Higgs with $m_{H^{\pm}}<89 \textnormal{GeV}$ has been excluded by LEP for all \tanb values \cite{lepexclusion1}. The Tevatron searches by D0 \cite{d01} and CDF \cite{cdf1} restrict the $m_{H^{\pm}}$ to be in the range $m_{H^{\pm}} > 80$ GeV for $2 <$ tan$\beta < 30$. The above results are followed by the indirect limit from $b \ra s\gamma$ studies by the CLEO collaboration which exclude $m_{H^{\pm}}$ below 300 GeV at 95 $\%$ C.L. in 2HDM Type II with \tanb$>2$ \cite{B1}. In general in terms of 2HDM types the current conclusion is $m_{H^{\pm}}>300$ GeV in 2HDM Type II and III. In addition to the above constraints, the CMS collaboration restricts a neutral MSSM Higgs boson to be heavier than 200 GeV with \tanb = 10, assuming $m_{h}$ boson around 125 GeV \cite{h2tautauCMS}. Therefore, an additional neutral Higgs should be heavier than at least 200 GeV with \tanb less than 10. Based on this observation, and to allow for high \tanb values, $m_{H^{0}}$ of 300 GeV is assumed in this study. New results from ATLAS indicate no charged Higgs lighter than 160 GeV with \tanb$>20$ \cite{atlconf2013}. Combined results from CMS search report \cite{cmschnew} and from ATLAS the fact that there is no exclusion for a $H^{\pm}$ heavier than 170 GeV. The chosen mass points of charged Higgs are also consistent with bounds from $B_s \rightarrow \mu^+\mu^-$ studies, as well as respecting $\Delta \rho$ parameter. Fig. \ref{brhbb}, which shows results for $m_{H}=300$ GeV, \tanb = 10 and with three adopted values of $m_{A}$. The global fit to SM electroweak measurements requires $\Delta \rho$ to be $O(10^{-3})$ \cite{pdg}. Since Fig. \ref{brhbb} shows $\Delta \rho$ values within the same order of magnitude, therefore, we conclude that the set of chosen mass points are consistent with EW precision measurements.\\

\section{Triple Higgs Couplings} 
The triple Higgs production can be regarded as a unique process for the charged Higgs studies as in that case, it contains a pair of charged Higgs and a single neutral Higgs ($H^+H^-H^0$ or $H^+H^-h^0$). Having the neutral Higgs decayed to $b\bar{b}$, the final state has effectively two extra $b-$jets as compared to the charged Higgs pair production. This feature makes it easy to be distinguished from the background processes. The reason is lack of existence of true $b-$jets in SM background events $WW$, $ZZ$ and $Z^{(*)}/\gamma^*$. The $t\bar{t}$ background can be reduced by a cut on the invariant mass of the two $b$-jets.\\
In what follows, the triple Higgs production is analyzed as the main source of charged Higgs bosons. To this end, the triple Higgs couplings are used for the signal production as presented in \cite{3H8,3H9,3H10,3H2}. We check briefly $H^0H^+H^-$ couplings presented in the mentioned references dealing one by one and showing in Eqn. (\ref{1}) only taken from \cite{3H8}.

\begin{eqnarray}
g_{H^+H^-H^0}=\frac{2m_W s_W}{e}&[&s_{\beta-\alpha}(\frac{1}{4} s_{2\beta}^2 (\lambda_1+\lambda_2) + \lambda_{345} (s_{\beta}^4+c_{\beta}^4) - \lambda_4-\lambda_5 - s_{2\beta} c_{2\beta}(\lambda_6-\lambda_7)) \nonumber \\  
&+& c_{\beta-\alpha} (\frac{1}{2} s_{2\beta} (s_{\beta}^2\lambda_1-c_{\beta}^2\lambda_2+c_{2\beta}\lambda_{345})-\lambda_6 s_{\beta} s_{3\beta}-\lambda_7 c_{\beta} c_{3\beta}))] 
\label{1}
\end{eqnarray}
With the assumptions $\lambda_5=\lambda_6=\lambda_7=0$ and by assuming $s_{\beta-\alpha}=1$ the above equation makes sure that the neutral lightest Higgs boson has the same couplings to gauge bosons as the SM partner. It is therefore an SM-like Higgs boson. This is due to the fact that the ratio of Higgs-gauge coupling in 2HDM to SM Higgs-gauge coupling can be expressed as follows \cite{3H11}:\\
\begin{equation}
\frac{g_{h_{2HDM}VV}}{g_{h_{SM}VV}}=s_{\beta-\alpha},~~\frac{g_{H_{2HDM}VV}}{g_{H_{SM}VV}}=c_{\beta-\alpha}
\end{equation}
Since we use large $\beta$ values, the above requirement ($s_{\beta-\alpha}=1$) leads to small and negative $\alpha$ values. Moreover, in this limit, the SM-like Higgs boson has the same coupling to a pair of bottom quarks as in SM, because \cite{3H11}\\
\begin{equation}
\frac{g_{h_{2HDM}b\bar{b}}}{g_{h_{SM}b\bar{b}}}=-s_{\alpha}/c_{\beta}=s_{\beta-\alpha}-t_{\beta}c_{\beta-\alpha}
\end{equation}
\section{Signal and Background Events and their Cross Sections}
\begin{figure}[htp]
\centering
\includegraphics[width=.4\textwidth]{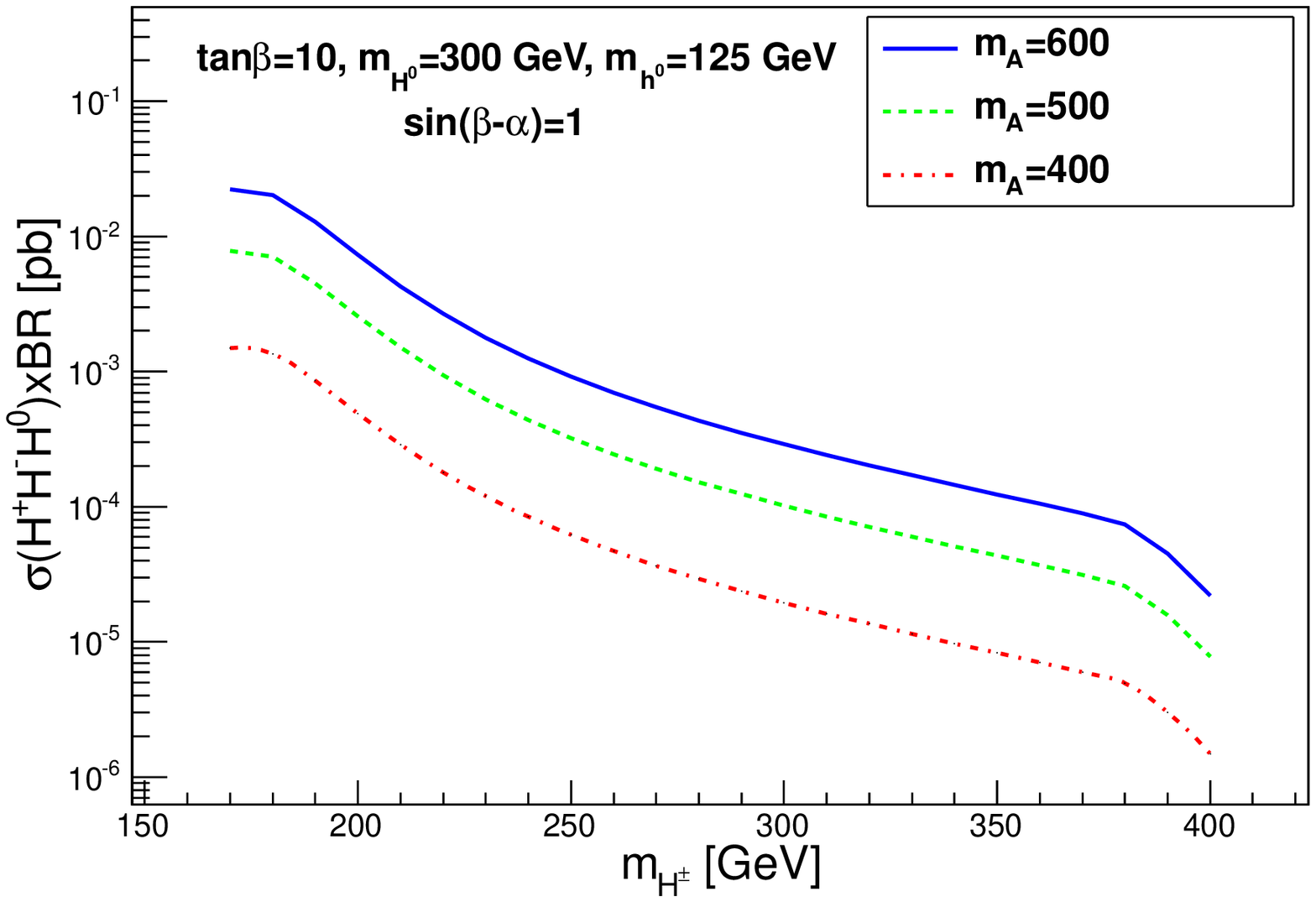}\hfill
\includegraphics[width=.4\textwidth]{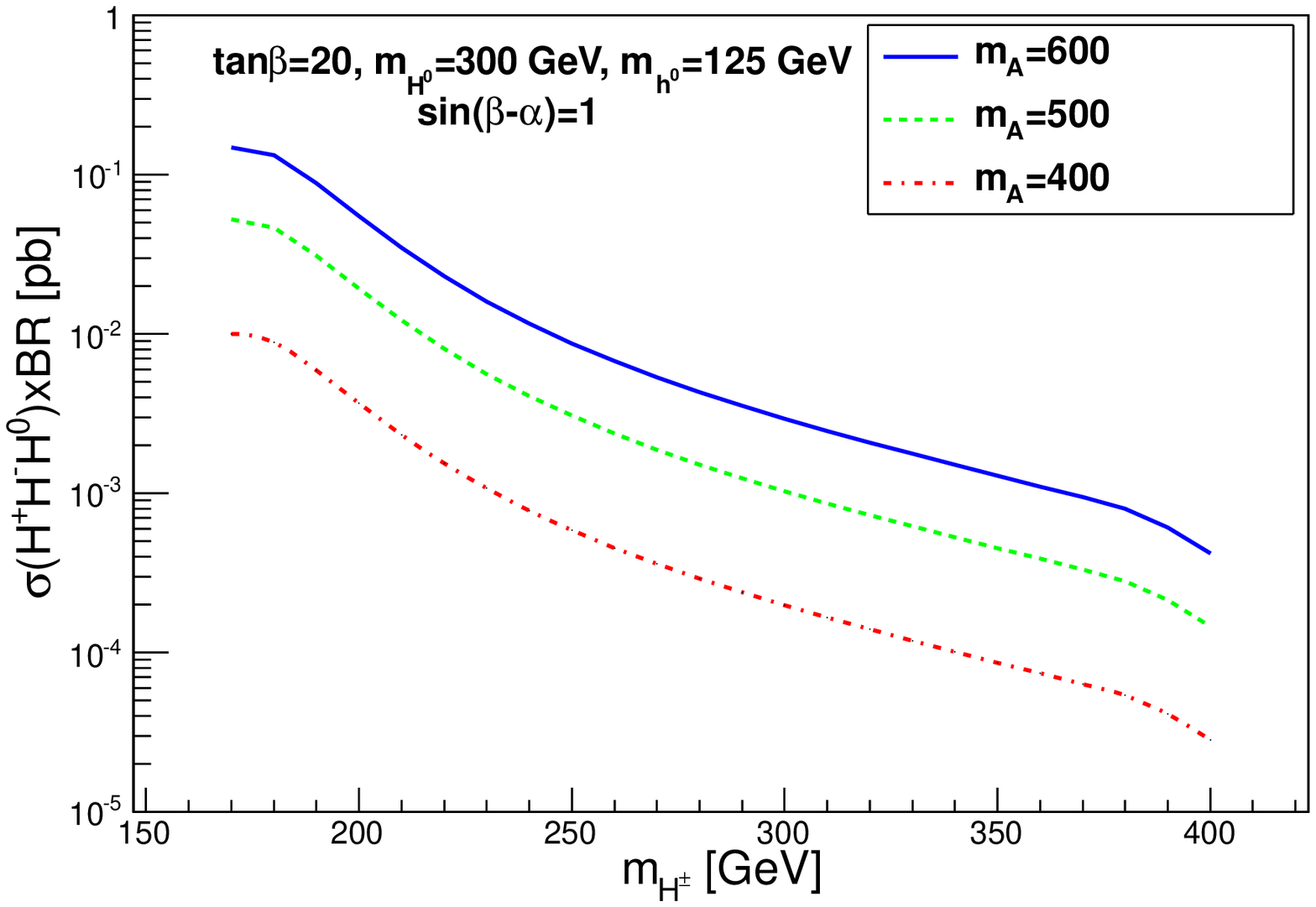}\hfill
\includegraphics[width=.4\textwidth]{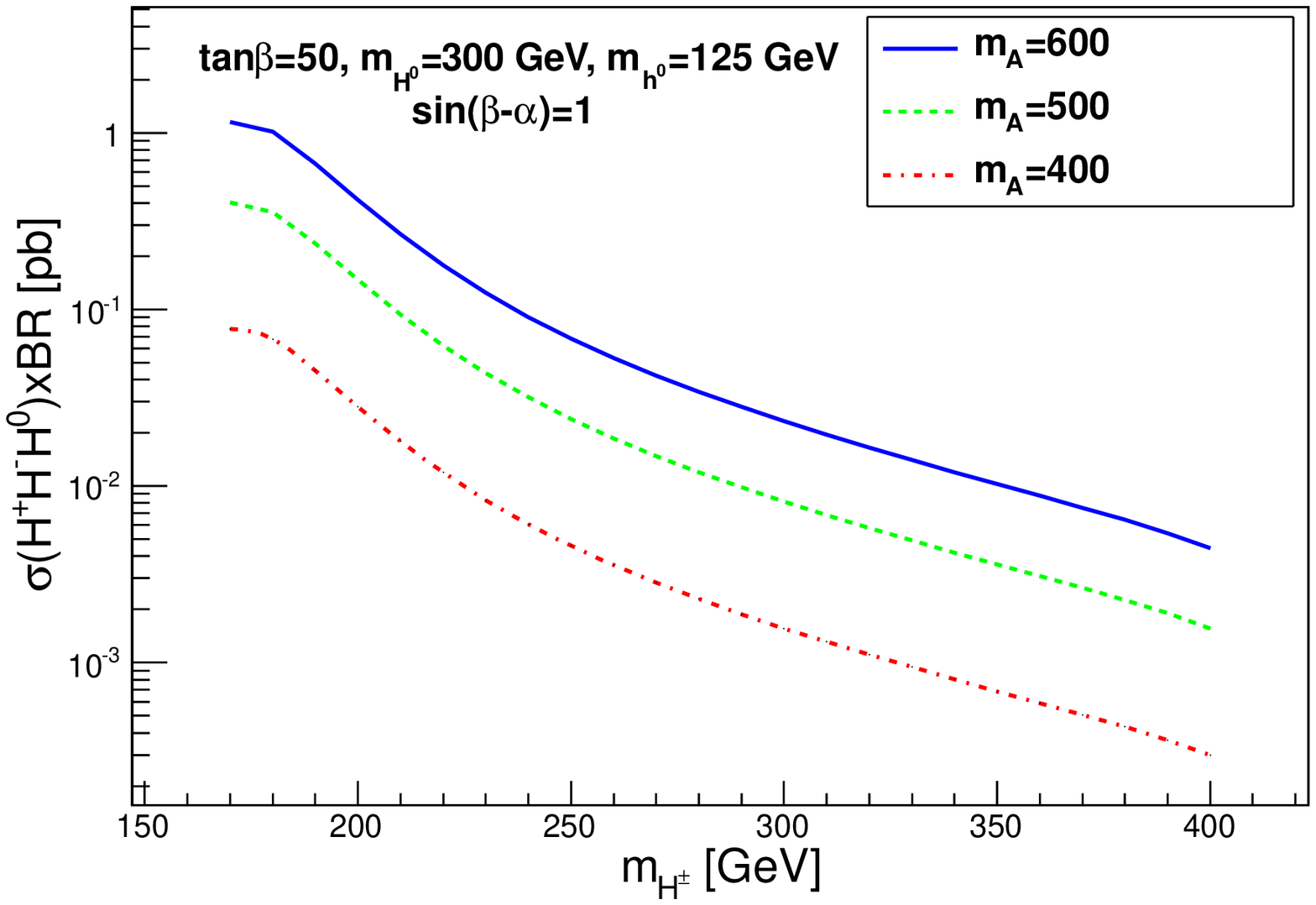}
\caption{The signal cross section times branching ratio of charged and neutral Higgs decays with \tanb=10, \tanb=20 and \tanb=50}
\label{sb1}
\end{figure}

The triple Higgs production can be either $H^+H^-H^0$ or $H^+H^-h^0$. However, according to the corresponding couplings presented in Eqns. (\ref{HHH}) and (\ref{HHh}) (Ref. \cite{3H2}), the $H^+H^-H^0$ coupling is larger than $H^+H^-h^0$ in the limit $s_{\beta-\alpha}=1$, unless there is a very large mass difference between the charged Higgs and CP-odd neutral Higgs.
\begin{equation}
H^{\pm}H^{\pm}H^0:~\frac{-ie}{m_W \sw s_{2\beta}}\left[(m_{H^{\pm}}^2-m_{A}^2+\frac{1}{2}m_{H}^2)s_{2\beta}\cbma-(m_{H}^2-m_{A}^2)c_{2\beta}\sbma \right] 
\label{HHH}
\end{equation} 
\begin{equation}
H^{\pm}H^{\pm}h^0:~\frac{-ie}{m_W \sw s_{2\beta}}\left[(m_{H^{\pm}}^2-m_{A}^2+\frac{1}{2}m_{h}^2)s_{2\beta} \sbma +(m_{h}^2-m_{A}^2) c_{2\beta}\cbma \right] 
\label{HHh}
\end{equation}
\begin{figure}[htp]
\centering
\includegraphics[width=.4\textwidth]{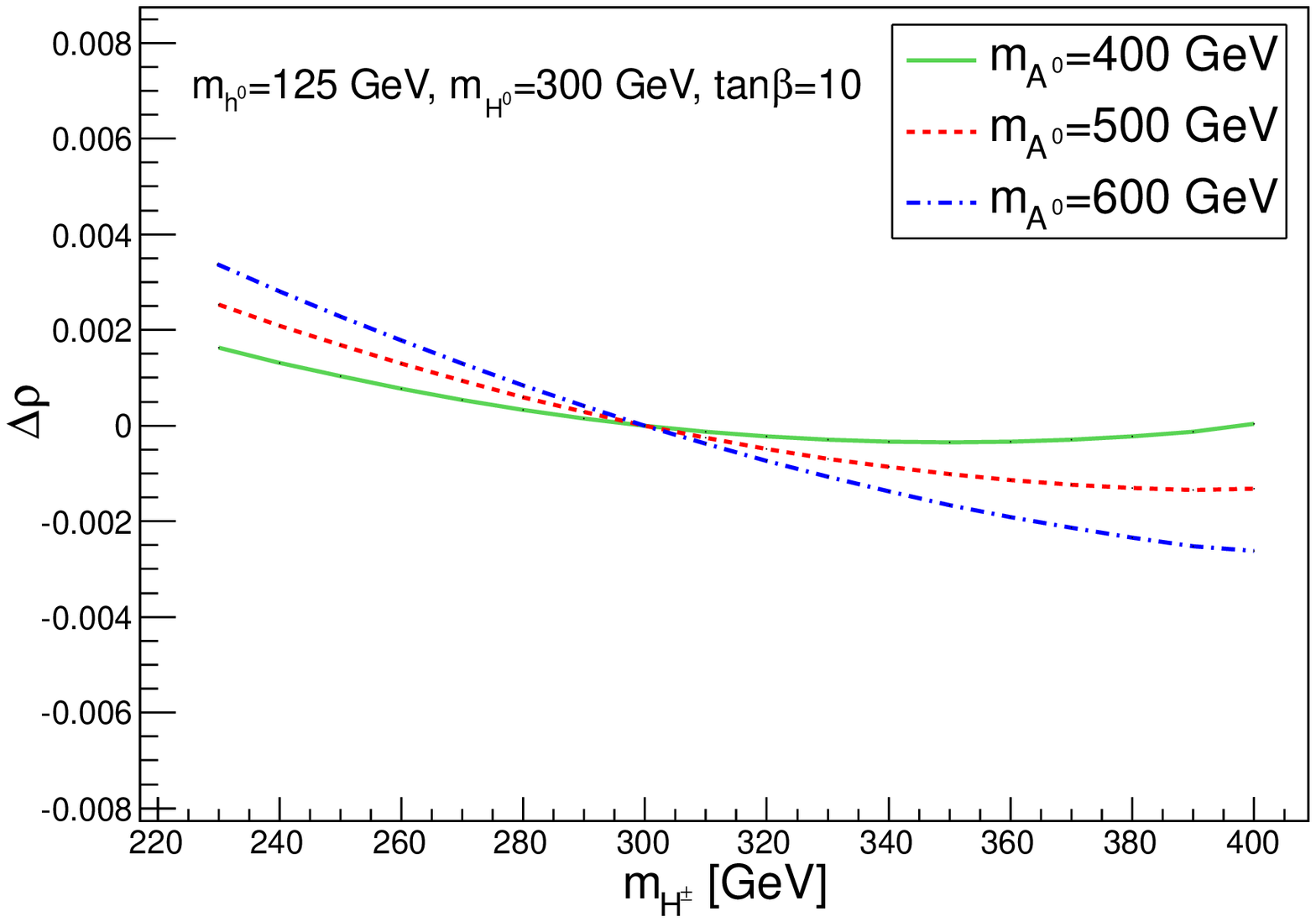}\hfill
\includegraphics[width=.4\textwidth]{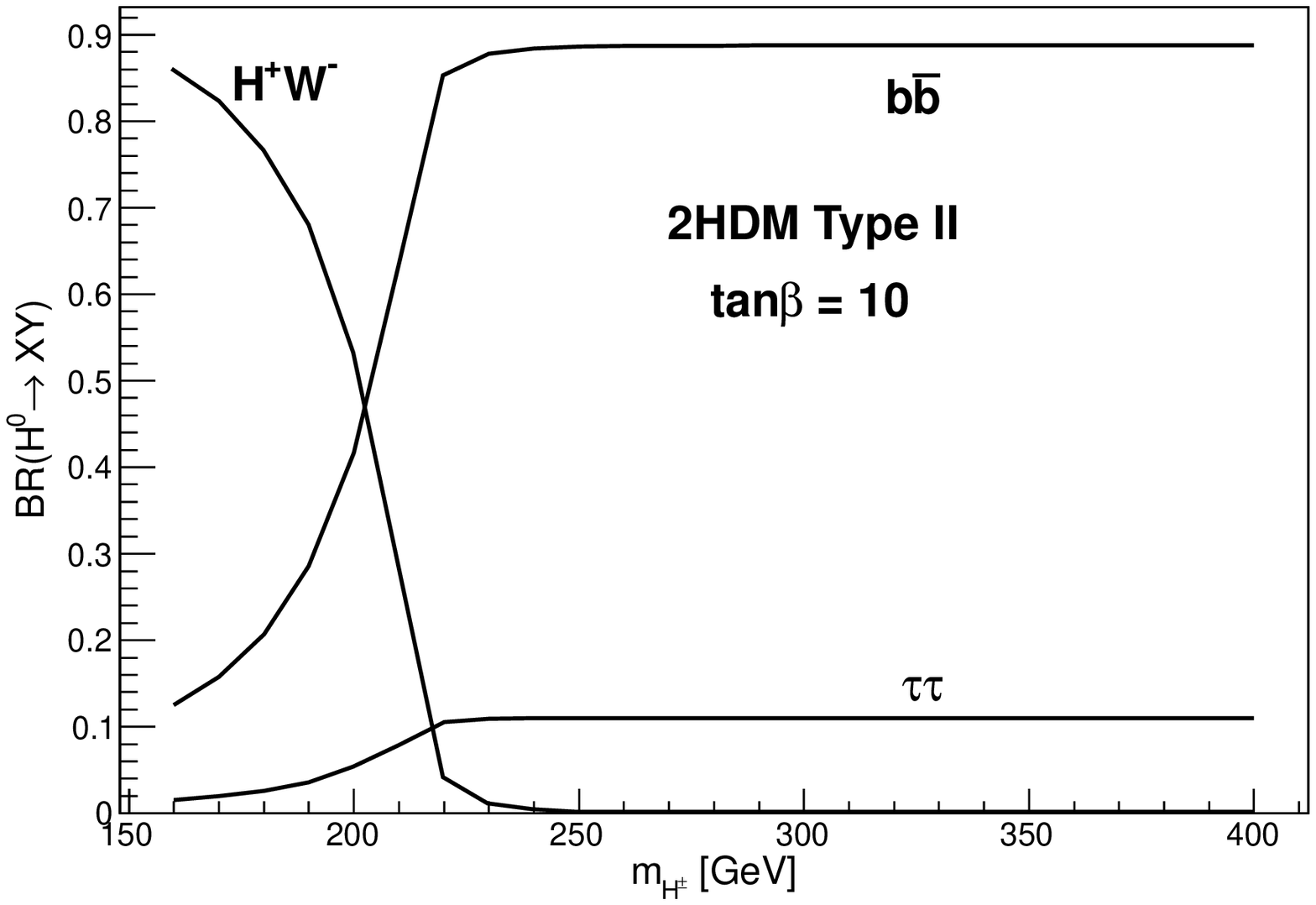}\hfill
\includegraphics[width=.4\textwidth]{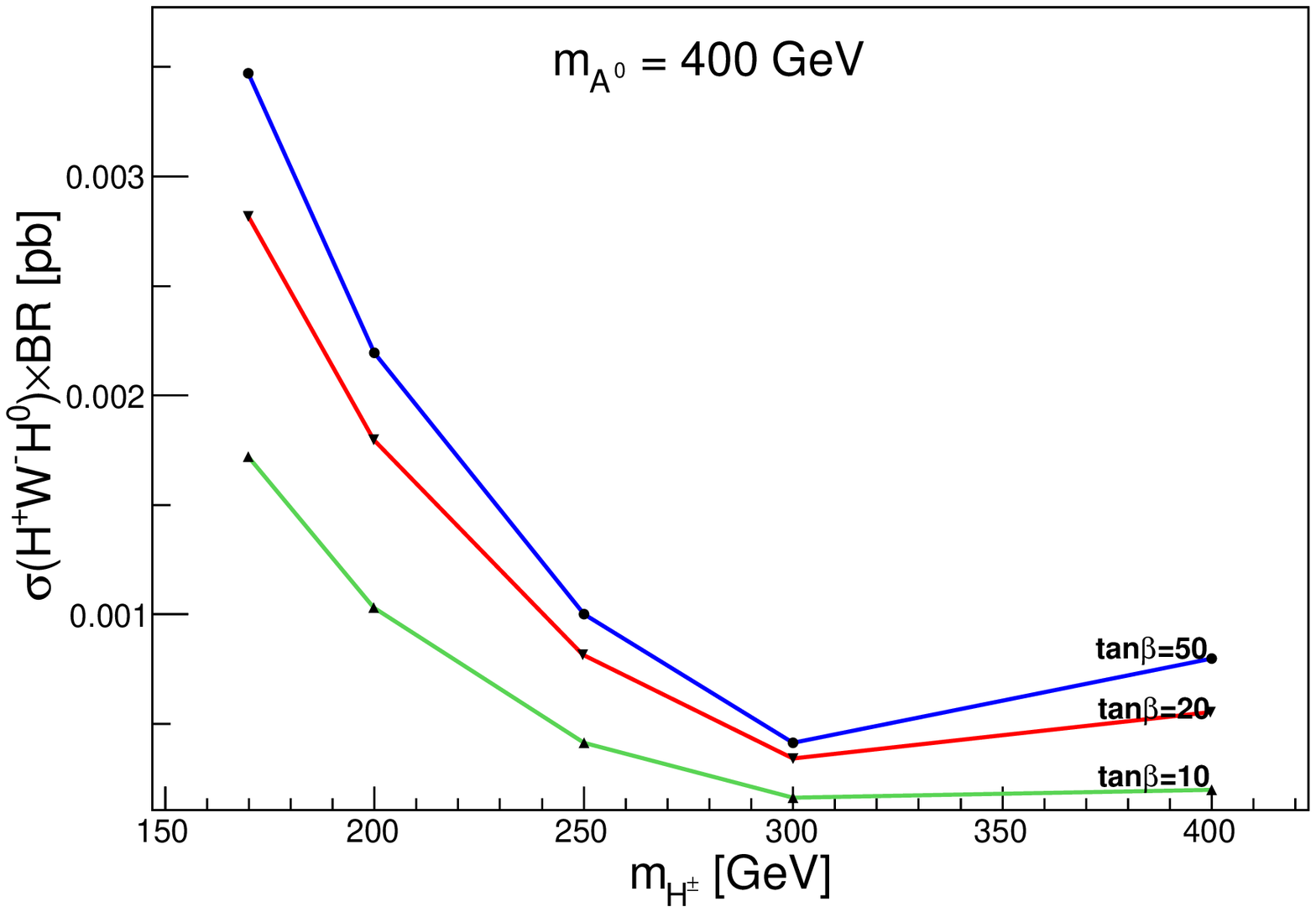}
\caption{(top left) $\Delta \rho$ parameter as a function of $m_{H^{\pm}}$ for different $m_{A}$ values. (top right), the neutral Higgs branching ratio of decays and (bottom), the $H^+W^-H^0$ signal cross section times branching ratio of charged and neutral Higgs decays.}
\label{brhbb}
\end{figure}
Therefore the signal is assumed to be $H^+H^-H^0$ or $H^+W^-H^0$ production. The latter process involves one charged Higgs but its cross section is comparable to the triple Higgs process. Therefore by signal we mean a sum of the above two processes. For reasonable background rejection, it is a convenient choice to assume the neutral Higgs decay to $b\bar{b}$. Since the charged Higgs decay to $t\bar{b}$ produces a high multiplicity event, it is better to choose $H^{\pm}\ra \tau \nu$ decays which produce low multiplicity events. According to the Higgs-fermion Yukawa couplings defined for the four types of 2HDM, the type-II 2HDM is most suitable for such a final state, because it provides the largest $H^{\pm} \ra \tau\nu$ and $H^0 \ra b \bar{b}$ branching ratio of decays at high \tanb. 
As a summary the full signal production process is  
\begin{equation}
e^+e^- \ra H^+H^-H^0 (H^+W^-H^0) \ra \tau^+ \tau^- b \bar{b} E^{\textnormal{miss}}_{T}
\end{equation}
The Higgs coupling depends on $m_A^2 - m_H^2$ according to Eq. (\ref{HHH}). The neutral Higgs mass ($m_{H^0}$) has to be greater than $m_{h^0}$, and be small enough to allow for heavy $m_{H^{\pm}}$ to be produced, however, it is constrained from below by LHC searches for neutral $H \ra \tau\tau$. A neutral Higgs with $m_{H^0}=300$ GeV is well outside the excluded area. Now the cross section can increase if $m_A$ increases resulting in larger $m_A^2 - m_H^2$ factors. Other decay channels such as $H^{+} \ra W^+H^0$ and $H^+ \ra W^+A^0$ lead to three or four particles for each charged Higgs decay and are not considered as their identification is difficult due to limited particle identification efficiencies and detector considerations. The main decay channels of the neutral Higgs are also shown in terms of branching ratios in Fig. \ref{brhbb}. Similar results are observed for other values of \tanb. The neutral Higgs branching ratio of decay to $b\bar{b}$ decreases with increasing Higgs boson mass, therefore higher cross sections are expected for lighter neutral Higgs bosons. The decay channel $H \ra H^{\pm}W^{\mp} $ leads to three charged Higgs bosons in the final state of the main process and is not considered here although it acquires a higher branchig ratio than $b\bar{b}$. Other decay channels, e.g., $H^{0} \ra h^{0}h^{0}$, $H^{0} \ra W^{+}W^{-}$ and $H^{0} \ra Z^{0}Z^{0}$ vanish as they are proportional to $c_{\beta-\alpha}$ and this analysis is based on $s_{\beta-\alpha}=1$. Therefore as long as the lightest neutral Higgs is required to be SM-like with $s_{\beta-\alpha}=1$, such decay channels do not play a role. The cross section times branching ratio of the triple Higgs production is thus obtained using branching ratio presented in \ref{brhbb}. Results are shown in Fig. \ref{sb1}. The double Higgs production originates from different decay chains, $e^+e^- \ra A^0 H^0 \ra H^+ W^- H^0$ and $e^+e^- \ra H^+ H^- \ra H^+ W^- H^0$. Therefore, its cross section depends on $m_{A^{0}}$, $m_{H^{0}}$ and $m_{H^{\pm}}$. Figure \ref{brhbb} shows the cross section of this process times BR$(H^{\pm}\ra \tau \nu)\times$BR$(H^{0}\ra b\bar{b})$. A double charged Higgs production through $e^+e^- \ra H^+ H^- \ra t \bar{b} \tau \bar{\nu} \ra W^+ b \bar{b} \tau \bar{\nu} \ra \tau^+ \nu b \bar{b} \tau \bar{\nu}$ was also explicitly checked and turned out to make no contribution to the signal as it was suppressed by the cut on the $b\bar{b}$ invariant mass.   

The background events are SM processes, $ZZ$, $WW$, $Z^{(*)}/\gamma^*$ and $t\bar{t}$ with cross sections 0.13, 1.8, 2.0 and 0.1  $pb$ respectively at $\sqrt{s}=1.5$ TeV. If $Z\ra \tau \tau$ and $Z \ra b\bar{b}$ decays occur, the $ZZ$ process can lead to $\tau \tau b \bar{b}$ final state. The $t\bar{t}$ events are also important background events as they contain two $b$-jets. Other sources of triple Higgs, e.g., $h^{0}H^+H^-$ and $A^{0}H^+H^-$ turn out to be negligible with a cross section of the order $10^{-6}$ pb. 
\section{Event Selection and Analysis}
\begin{figure}[htp]
\centering
\includegraphics[width=.4\textwidth]{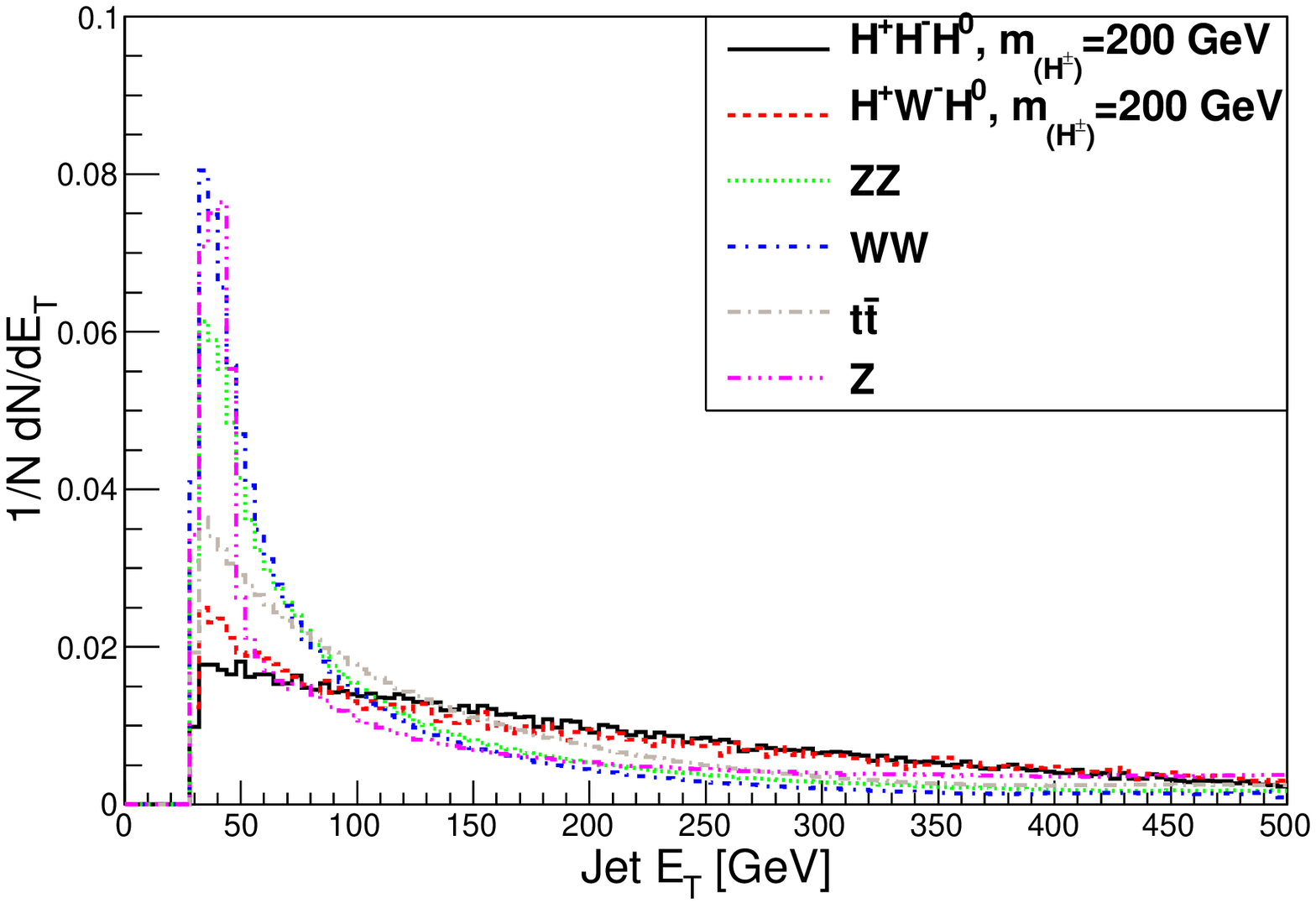}\hfill
\includegraphics[width=.4\textwidth]{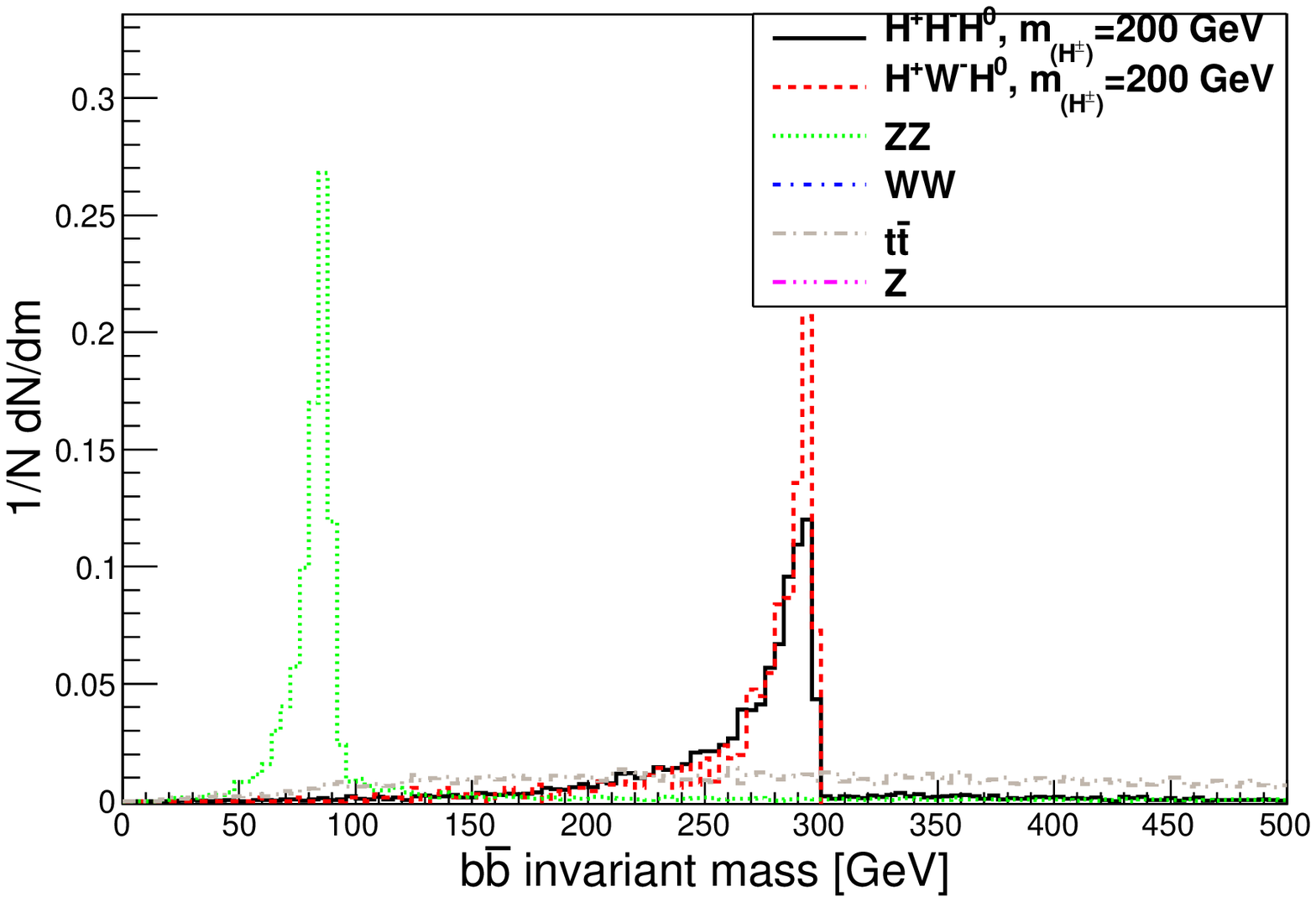}\hfill
\includegraphics[width=.4\textwidth]{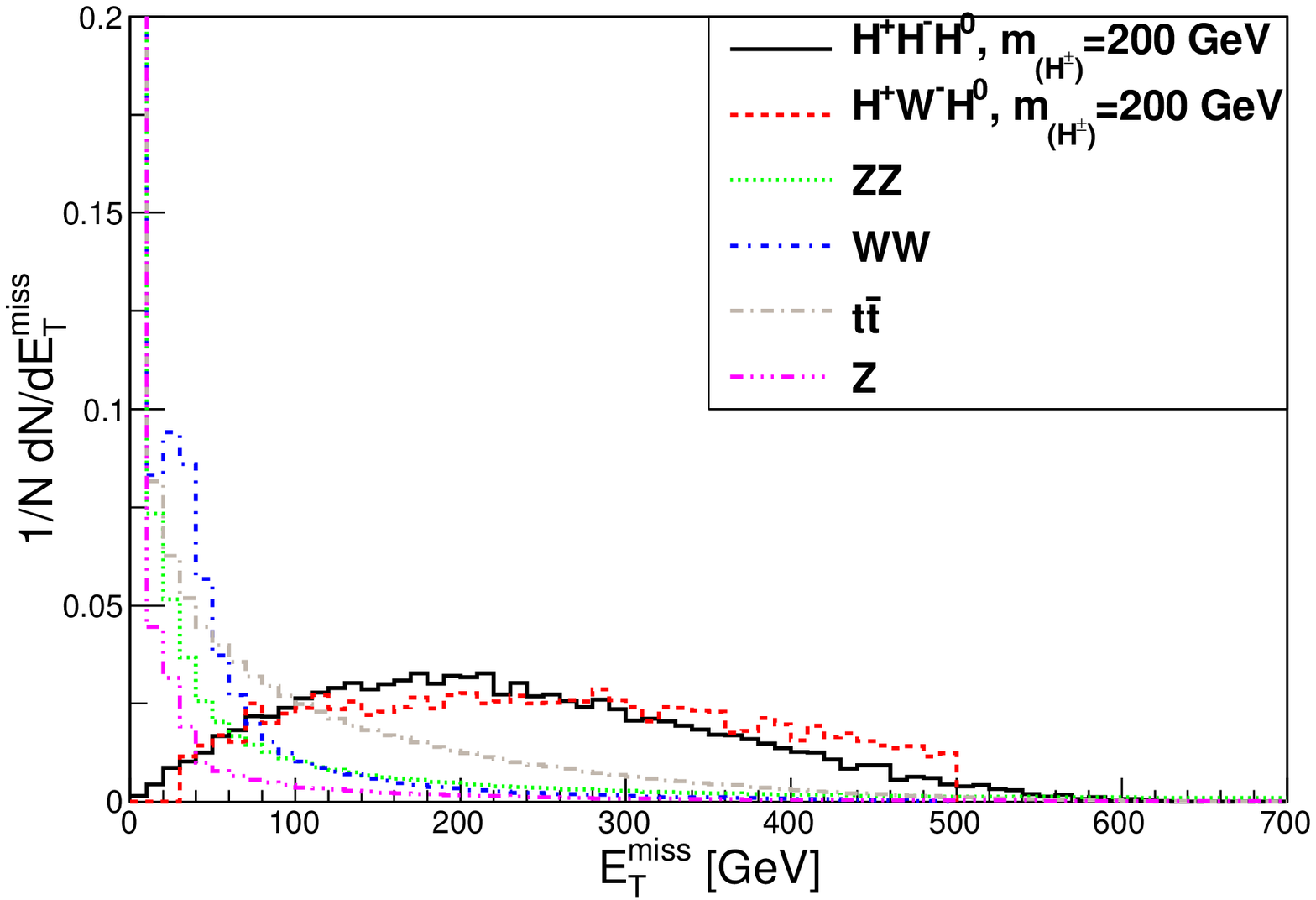}
\caption{(top left) the jet transverse energy distribution in signal and background events, (top right) the $b$-jet pair invariant mass distribution in signal and background events, while bottom figure is the missing transverse energy distribution in signal and background events.}
\label{jetet}
\end{figure}

Signal events contain four jets: two $\tau$-jets and two $b$-jets. Therefore three requirements on the number of jets are to be applied to separate signal and background, i.e., the cut on the number of all jets (with a kinematic cut as deduced from the jet transverse energy distributions), the cut on the number of $\tau$-jets which are signatures of the charged Higgs and a separated cut on the number of $b$-jets. The two $b$-jets originate from a neutral heavy Higgs boson in signal events, therefore, their invariant mass should lie within a mass window tuned by the neutral Higgs mass. Finally as there are two neutrinos in the event, a requirement on the minimum missing transverse energy should help suppression of some background events like single or double $Z$ bosons. From detector point of view, such requirements imply experimental uncertainties due to the jet energy scale, $b$-tagging efficiency, $\tau$-identification efficiency and missing transverse energy resolution.\\
In order to start event selection kinematic distributions are studied. Figure \ref{jetet} shows the (any) jet transverse energy distribution. Therefore the first step in signal selection is to require at least four jets in the final state with kinematic cuts on the jet transverse energy and pseudorapidity as in Eq. (\ref{jetetcut}). 
\begin{equation}
E^{\textnormal{jet}}_{T}~>~30 \textnormal{GeV},~~~~|\eta|<3
\label{jetetcut}
\end{equation} 
Selected jets are counted in the second step. The number of reconstructed jets passing the requirement of Eq. (\ref{jetetcut}). A cut on the number of reconstructed jets is applied as in Eq. (\ref{jetmulcut}).

\begin{equation}
\textnormal{Number of jets (satisfying Eq. (\ref{jetet})} \geq 4
\label{jetmulcut}
\end{equation}
The $\tau$-ID algorithm \cite{tauid} for $\tau$-jet reconstruction starts with a cut on the transverse energy of the hardest charged particle track in the $\tau$-jet cone as $E_{T} > 20 \textnormal{GeV}$. This requirement is basically applied as we expect a low charged particle multiplicity in $\tau$ hadronic decay which results in a large fraction of the $\tau$ energy to be carried by the leading track (the charged pion). The isolation requirement further uses the above feature of the $\tau$ hadronic decay by requiring no track with $p_T>1$ GeV to be in the annulus defined as $0.1< \Delta R <0.4$. Here $\Delta R=\sqrt{(\Delta \eta)^2 + (\Delta \phi)^2}$ and $\phi$ is the azimuthal angle. The $\Delta$ is calculated between the cone surface and the cone axis defined by the hardest track. The number of signal tracks are then calculated by searching for tracks in the cone defined around the hardest track with $\Delta R <0.07$. Since $\tau$ leptons decay predominantly to one or three charged pions, we require the number of signal tracks to be one or three. A jet (a $\tau$ lepton candidate) has to pass all above requirements to be selected as a $\tau$ lepton. Finally we require that there should be two $\tau$-jets satisfying all above requirements in the event.

In the next step the above two jets are used for $b\bar{b}$ invariant mass distribution as shown in Figure \ref{jetet} which tends to peak at 300 GeV (the neutral Higgs boson mass) in signal events while for the $ZZ$ background the $b\bar{b}$ invariant mass obviously peaks at the $Z$ mass. Based on this observation the requirement presented in Eq. (\ref{bbcut}) is applied on the distribution of $b$-jet pair invariant mass.  

\begin{equation}
b\bar{b} ~\textnormal{invariant mass} > 120 ~\textnormal{GeV}
\label{bbcut}
\end{equation}
As the last step, the missing transverse energy is reconstructed as the negative vectorial sum of particle momenta in the transverse plane as shown in Fig. \ref{jetet}. Based on the distribution shown in Fig. \ref{jetet}, the requirement of Eq. (\ref{metcut}) is applied on signal and background events.
\begin{equation}
E^{\textnormal{miss}}_T~>~30~ \textnormal{GeV}
\label{metcut}
\end{equation}

\section{Results}
\begin{figure}[htp]
\centering
\includegraphics[width=.4\textwidth]{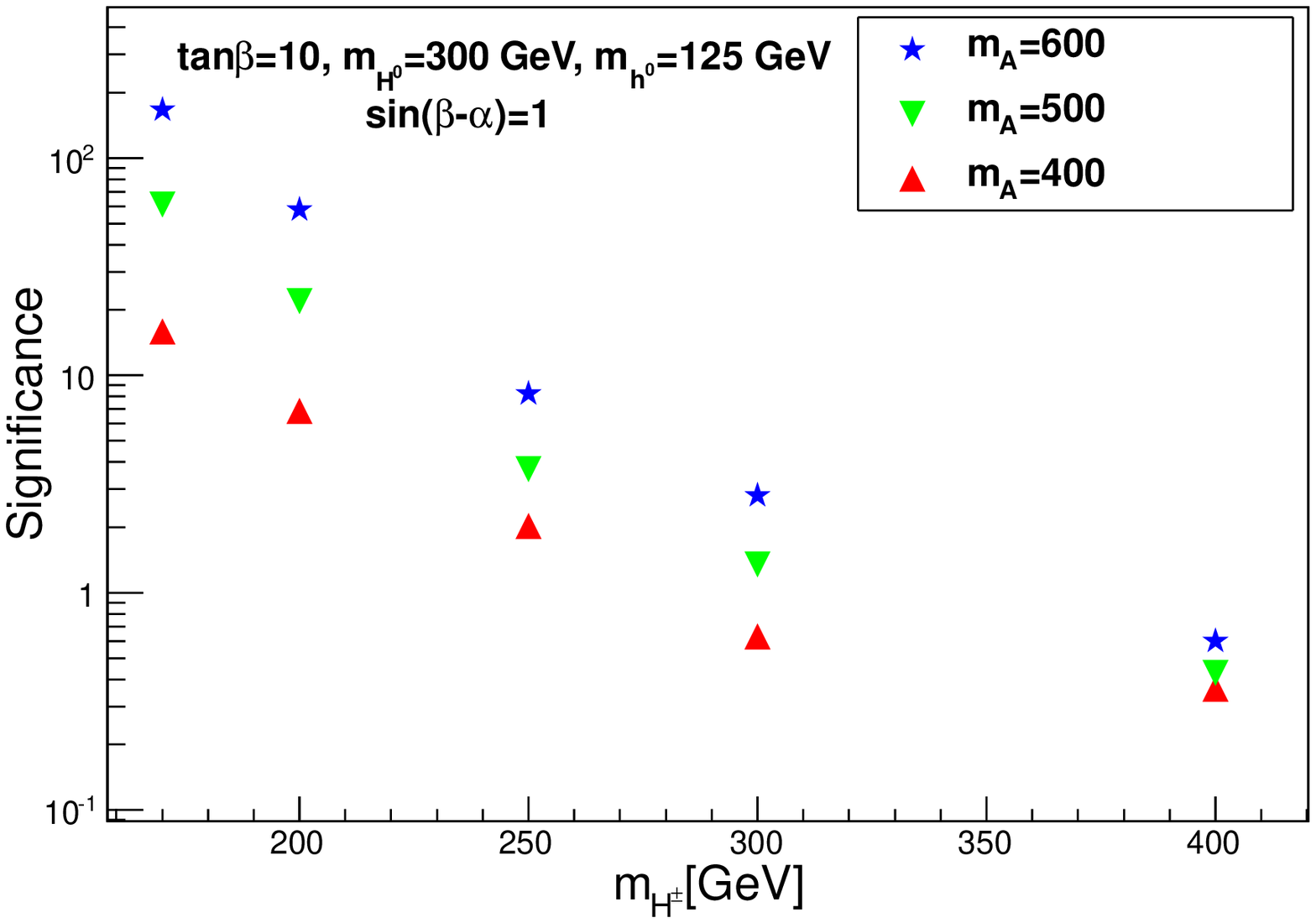}\hfill
\includegraphics[width=.4\textwidth]{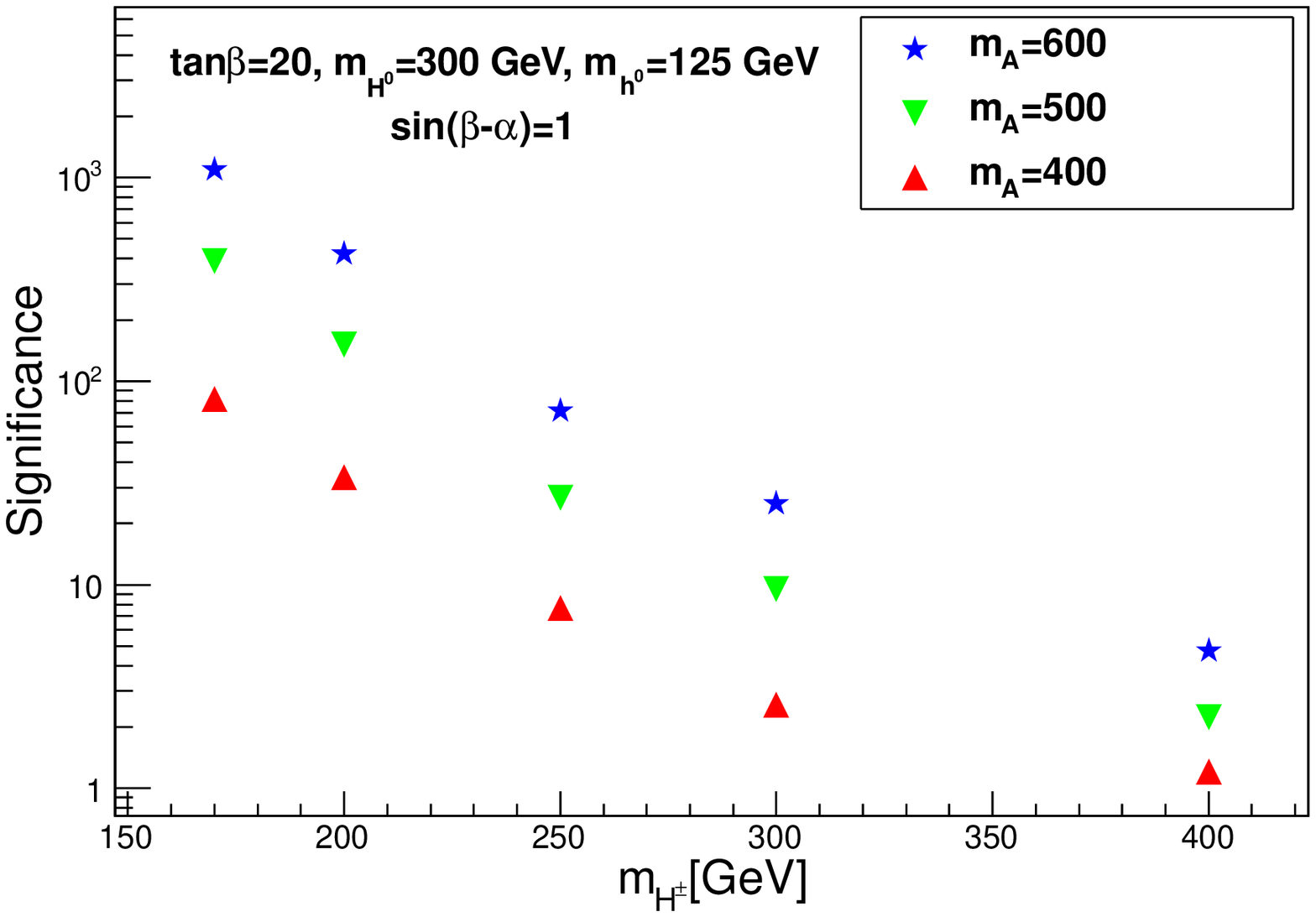}\hfill
\includegraphics[width=.4\textwidth]{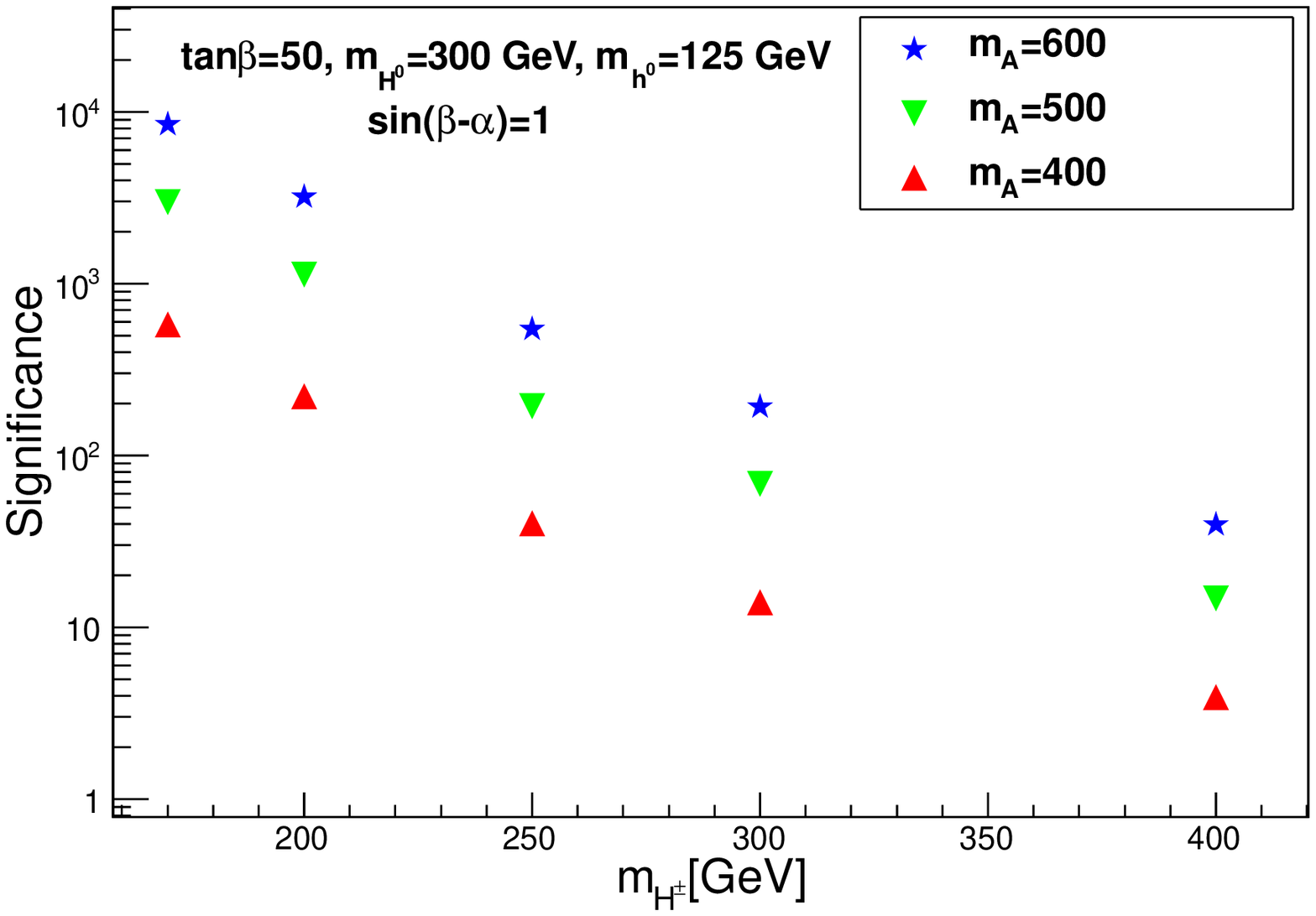}
\caption{The signal significance with \tanb=10, \tanb=20 and \tanb=50 a as a function of the $m_{H^{\pm}}$ and with different $m_A$ values at $\sqrt{s}=1.5$ TeV and integrated luminosity 500 $fb^{-1}$.}
\label{sig1}
\end{figure}

Selection cuts are applied one after the other, and relative efficiencies and the total efficiency of the signal and background selection is calculated. The final number of events, of course depends not only on the total selection efficiency, but also on the cross section of events. In case of signal, the cross section depends on \tanb, $m_{H^{\pm}}$ and $m_A$ and branching ratio of Higgs decays. Table \ref{seleff} shows the signal and background selection efficiencies.
\begin{table}[htbp]
\begin{center}
\tabcolsep 1.0pt
\small
\begin{tabular}{|c|c|c|c|c|c|c|c|c|c|c|c|c|}
\hline
& \multicolumn{8}{c|}{Signal, $m_{H^{\pm}}:$} & \multicolumn{4}{c|}{} \\
& \multicolumn{2}{c|}{170 GeV} & \multicolumn{2}{c|}{200 GeV} & \multicolumn{2}{c|}{300 GeV} & \multicolumn{2}{c|}{400 GeV} & \multicolumn{4}{c|}{Background} \\
\hline
&  HHH & HHW & HHH & HHW & HHH & HHW & HHH & HHW & ZZ & $Z^{(*)}/\gamma^*$ & WW & $t\bar{t}$ \\
\hline
Four jets & 0.64 & 0.64 & 0.63 & 0.63 & 0.63 & 0.63 & 0.62 & 0.62 & 0.24 & 0.052 & 0.22 & 0.91 \\
\hline
Leading track & 0.99 & 1 & 0.99 & 1 & 0.99 & 1 & 0.99 & 0.99 & 0.87 & 0.95 & 0.92 & 0.96 \\
\hline
Isolation & 0.87 & 0.69 & 0.88 & 0.69 & 0.9 & 0.63 & 0.9 & 0.83 & 0.35 & 0.092 & 0.6 & 0.24 \\
\hline
Number of signal tracks& 0.99 & 0.97 & 0.99 & 0.97 & 0.99 & 0.96 & 0.99 & 0.99 & 0.87 & 0.37 & 0.95 & 0.87 \\
\hline
Two $\tau$-jets & 0.41 & 0.2 & 0.43 & 0.21 & 0.47 & 0.22 & 0.5 & 0.1 & 0.68 & 0.6 & 0.19 & 0.071 \\
\hline
$\tau$-jet charge & 1 & 0.99 & 1 & 0.99 &  1 & 0.99 & 1 & 0.99 & 1 & 1 & 1 & 0.97 \\
\hline
Two $b$-jets & 0.82 & 0.84 & 0.82 & 0.84 & 0.82 & 0.82 & 0.82 & 0.81 & 0.15 & 0 & 0 & 0.61 \\
\hline
$b\bar{b}$ inv. mass & 0.88 & 0.94 & 0.88 & 0.93 & 0.87 & 0.93 & 0.85 & 0.96 & 0.041 & 0 & 0 & 0.33 \\
\hline
$E^{\textnormal{miss}}_{T}$ & 0.99 & 0.98 & 0.98 & 0.99 & 0.99 & 1 & 0.98 & 0.99 & 0.28 & 0 & 0 & 0.9 \\
\hline
Total eff. & 0.16 & 0.064 & 0.17 & 0.067 & 0.18 & 0.063 & 0.19 & 0.039 & $7.7\times 10^{-5}$ & 0 & 0 & 0.0023 \\
\hline
\end{tabular}
\end{center}
\caption{Signal and background selection efficiencies. HHH and HHW mean triple and double Higgs processes respectively. The signal selection efficiencies are assumed to be independent of \tanb and $m_A$. \label{seleff}}
\end{table}
The selection efficiencies are used in the next step to calculate the number of signal and background events at a given point in parameter space. The signal significance is calculated as $N_S/\sqrt{N_B}$, where $N_S(N_B)$ is the signal (background) number of events after all selection cuts. The significance depends on \tanb and $m_A$ due to the dependence of cross section to these parameters. Therefore different plots are produced for each value of \tanb and $m_A$ as shown in Fig. \ref{sig1}.

\section{Conclusion}
The triple Higgs boson production was analyzed as a source of charged Higgs pairs. The analysis was performed for a linear $e^+e^-$ collider operating at $\sqrt{s}=1.5$ TeV and results were presented with a normalization to an integrated luminosity of 500 $fb^{-1}$. The theoretical framework was set to 2HDM type-II containing an SM-like light Higgs boson with a mass equal to the current LHC observations. The effect of the CP-odd neutral Higgs mass in the production cross section and the signal significance was studied and it was concluded that increasing $m_A$ could increase the signal significance very sizably. The signal significance depends also on \tanb. A reasonable background suppression by b-tagging is achieved leading to high signal significance values for some areas of the parameter space. Finally the signal significance was presented as a function of the $m_{H^{\pm}}$, $m_A$ and \tanb. In such a scenario a linear collider with enough integrated luminosity higher than 500 $fb^{-1}$ would probably be the only experiment which could provide some news about this particle in the future in case the LHC missed to discover.

\end{document}